\documentclass[aps,preprint,showpacs,showkeys]{revtex4}
\usepackage[pdftex]{graphicx}
\usepackage{amssymb}
\usepackage{bm}
\usepackage{epsf,epsfig,subfigure,graphicx,amsmath,amssymb}

\newcommand\fverb{\setbox\fverbbox=\hbox\bgroup\verb}
\newcommand\fverbdo{\egroup\medskip\noindent%
			\fbox{\unhbox\fverbbox}\ }
\newcommand\fverbit{\egroup\item[\fbox{\unhbox\fverbbox}]}
\newbox\fverbbox

\newcommand{\mpl}{M_{\rm Pl}}

\newcommand{\del}{\partial}
\newcommand{\phie}{\phih_{\rm{end}}}
\newcommand{\phih}{\phi}
\newcommand{\gh}{g}
\newcommand{\lambdah}{\lambda}
\newcommand{\Mh}{M}

\begin{document}
\setcounter{page}{1}
\title[]{A Class of Inflation Models with Nonminimal Coupling}
\author{Seong Chan \surname{Park}}
\email{spark@phya.snu.ac.kr}
\affiliation{School of Physics and Astronomy, Seoul National University,\\
    Seoul 151-747, Korea}

\begin{abstract}
We show that the potential of the scalar field in the Einstein frame is flat if the non-minimal coupling term
is properly chosen so that it satisfies the condition ($V(\phi)/K^2(\phi) \rightarrow {\rm constant}$) as $\phi$ gets large. 
The cosmological implication of this theory is studied.
\end{abstract}


\maketitle

{\center \section{Introduction}}

In particle physics models, inflation takes place when a scalar field,
dubbed the inflaton field, can dominate the energy density of the universe ~\cite{Lyth:1998xn}(also see ~Ref.\cite{JKPS}). 
Under the slow-roll condition, the curvature perturbation is produced in a nearly scale-invariant way. The theory can be heavily constrained by the measurements of the anisotropies of the
Comic Microwave Background (CMB) and the observations of the large scale structure~\cite{obs}. The slow-roll
condition is that the inflaton potential must be very flat so that the effective
mass of the inflaton should be very small.  In model building,  the biggest question is the origin of the 
flatnees. That is what we want to find a clue to by considering the nonminimal coupling term.

Let us start with a theory that does not include the non-minimal coupling term. 
The theory is nothing but the original general relativity, which can be
nicely described by the Einstein-Hilbert action

\begin{eqnarray}
S_{\rm EH} = \int d^4 x \sqrt{-g} \left[-\frac{M^2}{2}R \right].
\end{eqnarray}
Here, the dimensionful parameter $M$ has been introduced to fit the overall dimension of the Einstein-Hilbert action.
The value of the mass parameter is precisely determined by the Newton constant: $M = 1/\sqrt{G}$. 

Now let us try to extend the theory by including a scalar field. For the moment, let us take the scalar field as a gauge singlet so that
we can write its kinetic and potential terms in a simple way:

\begin{eqnarray}
S_{\rm EH + scalar} = \int d^4 x \sqrt{-g} \left[-\frac{M^2}{2}R +{\cal L}_{\rm scalar}\right].
\end{eqnarray}
One thing is that, once we have a scalar field, the term $M^2$ has exactly the same mass dimension as the square of the scalar
field, $\phi^2$, so that we may add this term in the action without spoiling the spirit of the effective field theory: any term that is allowed by the symmetry of the theory has to be introduced if the term has the same mass dimensions as the other terms.
If the scalar is not a singlet field, but a multiplet field, the nonminimal term should read  $\phi \phi^*=|\phi|^2$.

\begin{eqnarray}
S_{\rm EH + scalar} = \int d^4 x \sqrt{-g} \left[-\frac{M^2 + a\phi^2}{2}R +{\cal L}_{\rm scalar}\right].
\end{eqnarray}

This is the theory that we want to consider here \cite{Park and Yamaguchi, Park:SUSY08}. 

The goal of the study is to clarify the issue of the recent paper by Bezrukov and Shaposhnikov (BS) ~\cite{Bezrukov:2007ep} where the authors reported the very interesting possibility that the standard model Higgs field with the nonminimal term can give rise to  successful inflation~\cite{inf, books}. We extend the possibility by generalizing the form of the nonminimal coupling term by introducing a generic function $K(\phi)$.  Actually, the models of chaotic inflation with nonzero $a$ have been considered in various different contexts
\cite{Spokoiny:1984bd,Salopek:1988qh,Kaiser:1994vs,Komatsu:1999mt,Futamase:1987ua,Fakir:1990eg,Libanov:1998wg}.
The authors showed that the Higgs potential, which should be defined in Einstein frame, is nearly flat in the large-field limit.
Here, the authors required from the COBE data $U/\varepsilon=(0.027 \mpl)^4$ that the ratio
between the quartic coupling of the Higgs field ($\lambda$) and the nonminimal coupling constant ($a$)
should be small, $\sqrt{\lambda/a^2}\sim 10^{-5}$.

In the next section, after reviewing the idea of BS, we generalize the idea
by generalizing the form of the gravity-scalar coupling term in a nonminimal way ($K(\phi)R$). Then, we will read the general
condition for getting the flat potential or the slow-roll condition. In Sec. ~\ref{monomial},
we work out the monomial case with the functions $K\sim a \phi^m$ and  $V\sim \lambda \phi^{2m}$ in detail.

\vspace{0.5cm}
{\center \section{The Higgs boson as the inflaton and its generalization}}
\label{review}

Let us review the main idea of BS. BS consider a theory in the Jordan frame as follows:
\begin{eqnarray}
S_J = \int d^4 x \sqrt{-g} \left[-\frac{M^2+ a \phi^2}{2}R + \frac{(\partial \phi)^2}{2}-V(\phi)\right].
\end{eqnarray}
The scalar potential is $V(\phi) = \lambda/4 (\phi^2 -v^2)^2$ so that the electroweak symmetry breaking
is triggered by the nonzero vacuum expectation of the Higgs field. $M\simeq \mpl$ is given in the parameter region
$1\ll a \ll (\mpl/\langle\phi\rangle)^2$. They showed that one can transform to the Einstein frame in which the graviton
kinetic term is canonical. The transformation is conformal:
\begin{eqnarray}
g_{\mu\nu}\rightarrow g^E_{\mu\nu} = e^{2 \omega} g_{\mu\nu},
\end{eqnarray}
where $g^E$ is the metric in the Einstein frame and the conformal factor is defined as $e^{2\omega} = 1+ a \phi^2/\mpl$.
The action in the Einstein frame provides the physical Higgs potential
\begin{eqnarray}
U(h(\phi)) = e^{-4 \omega} V(\phi)
\label{potential}
\end{eqnarray}
where the canonically normalized scalar field $h$ is defined by the derivative
\begin{eqnarray}
\frac{d h }{ d\phi} = \sqrt{\frac{e^{2\omega}+ 6 a^2 \phi^2/\mpl^2}{e^{4 \omega}}}.
\end{eqnarray}
The numerical coefficient `6' comes from the conformal transformation of the Ricci scalar, $R\rightarrow e^{-2\omega}(R-6 \nabla^2 \omega - 6(\partial \omega)^2)$. In $D$ dimensions, $R\rightarrow e^{-2\omega}\left(R- 2(D-1)\nabla^2\omega-(D-2)(D-1)(\partial \omega)^2\right)$ in general.
Now let us take a "large field"  $h \gg \mpl/\sqrt{a}$, the potential in eq.\ref{potential} can be recasted as
\begin{eqnarray}
U(h) &\simeq& \frac{\lambda \mpl^4}{4 a^2}\left(1+ e^{-2 h/(\sqrt{6}\mpl)}\right)^{-2} \nonumber \\
&\simeq&  \frac{\lambda \mpl^4 }{4 a^2}.
\label{potential in Einstein}
\end{eqnarray}
It is clear that the potential becomes very flat in the large-field limit. BS found that if the coefficient of
the nonminimal coupling, $a$, is tuned, this potential really reproduces the current observational data. 
The fitted value is $\sqrt{\lambda/a^2} = 2.1 \times 10^{-5}$. As seen in Eq. \ref{potential in Einstein},
the flat potential is not directly related to the actual vacuum expectation value at low energy. Indeed, inflation
takes place in the large-field region, which should be independent of the details of the low-energy values as with the usual
effective field theories.

Now, let us generalize the model with nonminimal coupling $K(\phih)$ and the scalar potential $V(\phih)$. 
The action in Jordan frame is given as
\begin{eqnarray}
 S=\int d^4 x \sqrt{-\gh}\left[-\frac{\Mh^2+K(\phih)}{2}{R}+\frac{1}{2}(\del \phih)^2-V(\phih)\right].
\end{eqnarray}
Taking the Einstein metric,
\begin{eqnarray}
 \gh_{\mu\nu}=e^{-2\omega}g^{E}_{\mu\nu},\qquad
 e^{2\omega}:=\frac{\Mh^2+K(\phih)}{\mpl^2},
\end{eqnarray}
the action in the Einstein frame becomes
\begin{eqnarray}
 \int d^4x\sqrt{-g_{E}}\left[
-\frac{\mpl^2}{2}R_{E}
+\frac34\frac{e^{-4\omega}}{\mpl^2}K'(\phih)^2(\del \phih)^2
+\frac12 e^{-2\omega}(\del \phih)^2
-e^{-4\omega}V(\phih)
\right].
\end{eqnarray}
We should redefine the scalar field in such a way that the field is canonically  normalized in the physical frame.
\begin{eqnarray}
 \frac{dh}{d\phih}=\sqrt{
\frac{\mpl^2}{\Mh^2+K(\phih)}
+\frac32\frac{\mpl^2}{(\Mh^2+K(\phih))^2}K'(\phih)^2
}.\label{dhdphi}
\end{eqnarray}
Finally, the scalar potential is given as
\begin{eqnarray}
 U=\frac{\mpl^4}{(\Mh^2+K(\phih))^2}V(\phih). \label{U}
\end{eqnarray}
The potential will be flat if the condition 
\begin{eqnarray}
\lim_{\phih\rightarrow \infty}\frac{V}{K^2} = Const >0
\label{condition}
\end{eqnarray}
is satisfied because $U \varpropto \frac{V}{K^2}$. The condition $K(\phih) \gg \Mh^2$ for $\phih \gg \Mh$
is required for the potential to be bounded from below, and the location of the global minimum
is well localized around the small field value.

\vspace{0.5cm}
{\center \section{Monomial case: $K\sim \phi^m$}\label{monomial}}

The simple case we can think of is a monomial function.
Let $K(\phih)$ be a monomial as
\begin{eqnarray}
 K(\phih)=a \phih^m,
\end{eqnarray}
where $a$ is a dimensionful constant because $[K]=2$.
In order to get the flat potential in large $\phih$ region in the Einstein frame, 
the original scalar potential in the Jordan frame should be written as
\begin{eqnarray}
 V=\frac{\lambdah}{2m}\phih^{2m}.
\end{eqnarray}
In this case, $U$ is written as
\begin{eqnarray}
 U=\frac{\mpl^4\lambdah}{2m a^2}\left(1+\frac{\Mh^2}{a}\phih^{-m}\right)^{-2}
 \label{potentialm}
\end{eqnarray}

In the large-$\phih$ region, the relation in Eq. \ref{dhdphi} between $\phih$ and $h$ is written as
follows:
\begin{itemize}
 \item $m=1$
\begin{eqnarray}
 \frac{dh}{d\phih}\cong\frac{\mpl}{\sqrt{a}}\frac{1}{\sqrt{\phih}},\qquad
 \phih\cong \frac{a}{4\mpl^2}h^2. \label{hphi1}
\end{eqnarray}
 \item $m=2$
\begin{eqnarray}
 \frac{dh}{d\phih}\cong\sqrt{6+1/a}\frac{\mpl}{\phih},\qquad
 \phih\cong \frac{\mpl}{\sqrt{a}}\exp\frac{h}{\sqrt{6+1/a}\mpl}.\label{hphi2}
\end{eqnarray}
 \item $m\ge 3$
\begin{eqnarray}
 \frac{dh}{d\phih}\cong\sqrt{\frac32}\frac{m \mpl}{\phih},\qquad
 \phih\cong \left(\frac{\mpl^2}{a}\right)^{1/m}\exp\sqrt{\frac23}\frac{h}{m \mpl}.\label{hphi3}
\end{eqnarray}
\end{itemize}
Now, we are ready to consider the cosmological implications of the model. 

The slow-roll parameters are defined by using the scalar potential in the Einstein frame, Eq. \ref{U}, and the canonically normalized scalar field $h$ as
\begin{eqnarray}
 \varepsilon=\frac{\mpl^2}{2}\left(\frac{\del U/\del h}{U}\right)^2,
 \qquad
 \eta=\mpl^2\frac{\del^2 U/\del h^2}{U}.
\end{eqnarray}
In our model, these parameters are calculated in the large-$\phih$ region by using eqs.
\ref{potentialm} and \ref{hphi1}--\ref{hphi3} as
\begin{eqnarray}
 \varepsilon=
 \left\{
   \begin{array}{ll}
     \frac{2M}{a}\left(\frac{M}{\phih}\right)^3, & \hbox{$m=1$;} \\
     \frac{4}{3a^2(1+1/(6a))}\left(\frac{M}{\phih}\right)^4, & \hbox{$m=2$;} \\
     \frac{4M^{-2m+4}}{3a^2}\left(\frac{M}{\phih}\right)^{2m}, & \hbox{$m\ge 3$.}
   \end{array}
 \right.
,
 \eta=
 \left\{
   \begin{array}{ll}
     -3\left(\frac{\Mh}{\phih}\right)^2, & \hbox{$m=1$;} \\
     -\frac{4}{3a(1+1/(6a))}\left(\frac{\Mh}{\phih}\right)^2, & \hbox{$m=2$;} \\
     -\frac{4\Mh^{2-m}}{3a}\left(\frac{\Mh}{\phih}\right)^m, & \hbox{$m\ge 3$.}
   \end{array}
 \right.
\label{epsilon-eta}
\end{eqnarray}

The end of inflation is fixed by the condition $\varepsilon=1$. 
The values of $h$ and $\phih$ at this point are denoted by $h_{\rm end}$ and $\phie$ respectively. 
In the slow-roll inflation, the number of e-foldings is expressed as
\begin{eqnarray}
 N=\frac{1}{\mpl^2}\int^{h_0}_{h_{\rm{end}}}\frac{U}{\partial U/\partial h}.
\end{eqnarray}
In our model, $N$ is calculated as
\begin{eqnarray}
 N=
\left\{
\begin{array}{ll}
 \frac{1}{4M^2}(\phih_0^2-\phie^2), & \hbox{$(m=1)$}\\
 \frac{3}{4} a\left(1+\frac{1}{6a}\right)\frac{1}{M^2}(\phih_0^2-\phie^2), & \hbox{$(m=2)$}\\
 \frac{3}{4} a\frac{1}{\Mh^{2}}(\phih_0^m-\phie^m) , & \hbox{$(m \ge 3)$}
\end{array}
\right.
\end{eqnarray}
In order to get $60$ e-foldings, we should solve $N=60$ and get $\phih_{60}$.
Let us assume $\phih_{60} \gg \phie^2$. Then, we obtain the value $\phih_{60}$ as
\begin{eqnarray}
  \phih_{60}=
\left\{
  \begin{array}{ll}
    2\sqrt{N}M, & \hbox{($m=1$)} \\
    \frac{2 \sqrt{N}M}{\sqrt{3a(1+1/(6a))}}, & \hbox{($m=2$)} \\
    \left(\frac{4 N}{3a}M^2\right)^{1/m}, & \hbox{($m\ge 3$).}
  \end{array}
\right.
\label{phi60}
\end{eqnarray}

The spectral index $n_s$ and the tensor-to-scalar ratio $r$ can be calculated as
\begin{eqnarray}
 n_s=1-6\varepsilon+2\eta|_{\phih=\phih_{60}},\qquad
 r=16\varepsilon|_{\phih=\phih_{60}}.
\end{eqnarray}
In our model, these values are expressed (using Eq.~ \ref{epsilon-eta} and Eq. ~\ref{phi60}) as
\begin{eqnarray}
 n_s=
\left\{
\begin{array}{ll}
 1-\frac{3}{2a_0N^{3/2}}-\frac{3}{2N}, & \hbox{$(m=1)$}\\
 1-\frac{9(1+1/(6a_0))}{2N^2}-\frac{2}{N} & \hbox{$(m=2)$}\\
 1-\frac{9}{2N^2}-\frac{2}{N}, & \hbox{$(m\ge 3),$}
\end{array}\right.
\qquad
r=
\left\{
\begin{array}{ll}
 \frac{4}{a_0N^{3/2}}, & \hbox{$(m=1)$}\\
 \frac{12(1+1/(6a_0))}{N^2} & \hbox{$(m=2)$}\\
 \frac{12}{N^2}  & \hbox{$(m\ge 3)$,}
\end{array}\right.
\end{eqnarray}
where the dimensionless parameter $a_0$ is defined as
\begin{eqnarray}
 a_0=a M^{m-2}.
\end{eqnarray}

\begin{figure}[t!]
\includegraphics[width=13.0cm]{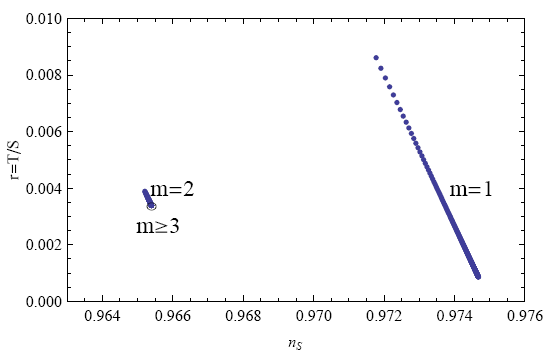}
\caption{The spectral index $n_s$ and the tensor-to-scalar perturbation ratio $r$
are depicted with the various values of $a_0$ and the power of the
non-minimal coupling $m$ in $K(\phi)\sim \phi^m$. }
\label{rnplot}
\end{figure}

In Fig.~\ref{rnplot}, we plotted the spectral index ($n_S$) and the tensor-to-scalar
perturbation ratio ($r$) for values of $a_0$. The number of e-foldings is fixed at $N=60$. 
For the monomial cases with $m=1$ and $m=2$,
the spectral index becomes larger but the tensor-to-scalar ratio becomes smaller.
For {\it large} $a_0\simeq 4\pi$,
the values of the spectral index and the tensor-to-scalar ratio are saturated to
$0.9745 (0.965)$ and $0.0007(0.003)$ for $m=1(m\geq 2)$, respectively. Notice that
when $m\geq 3$, the spectral index and $r$ are independent of $a_0$ and
are given as $0.965$ and $0.003$, respectively. This corresponds to the circle at the tip
of the plot for $m=2$.

The amplitude of the scalar perturbation is another nice observable:
\begin{eqnarray}
 \delta_{H}=\frac{\delta \rho}{\rho}\cong \frac{1}{5\sqrt{3}H}\frac{U^{3/2}}{\mpl U'}=1.91\times 10^{-5}.
\end{eqnarray}
This gives a constraint for the parameters
\begin{eqnarray}
 \frac{U}{\epsilon}=(0.027\mpl)^4.
\end{eqnarray}
In our model, the constraint is written, with the dimensionless parameter
 $\lambda_0=\lambda \Mh^{2m-4}$, as follows.
\begin{eqnarray}
 \begin{array}{ll}
  \sqrt{\frac{\lambda_0}{a_0}}&\simeq 2.3\times 10^{-5},~~ (m=1)\\
  \sqrt{\frac{\lambda_0}{a_0^2(1+1/(6a_0))}}&\simeq 2.1\times 10^{-5},~~ (m=2)\\
  \sqrt{\frac{\lambda_0}{a_0^2}}&\simeq 1.5\times 10^{-5}\sqrt{m},~~ (m\ge 3).\\
 \end{array}
\label{condition2}
\end{eqnarray}
One should notice that the smallness of the values
can be understood in terms of compactification once the theory is
embedded in higher dimensions. This will be discussed in a new paper in the future.
One should note that the condition $\sqrt{\frac{\lambda_0}{a_0^2}}\sim 10^{-5}$ is
universally required to fit the observational data.

\vspace{0.5cm}
{\center \section{Conclusion}}

We studied  inflationary scenarios based on nonminimal coupling of a
scalar field. Taking a conformal transformation, the scalar potential in the Einstein frame 
is found to be flat in the large-field limit if the condition
in Eq.~\ref{condition} is satisfied. This is the main result of this paper. This model is constrained by cosmological observations.
The spectral index, the tensor-to-scalar perturbation ratio, and the amplitude of the potential are nice 
observable quantities for us to be able to compare the theory with the data. 
We considered the case with a monomial function $K \sim \phi^m$ and showed that this class of models is in good
agreement with the recent observational data: $n_S \simeq 0.964-0.975$ and $r \simeq 0.0007- 0.008$ for any value of $m$.
In Fig.~\ref{rnplot}, the predicted values for $n_S$ and $r$ are given  in the observational bound. We read out the condition for
fitting the observed anisotropy of the CMBR by which essentially the amplitude of the potential is determined. The condition
does not look natural ($\sqrt{\lambda/a^2}\sim 10^{-5}$) at  first sight, but we may understand this seemingly unnatural value
once we embed the theory in higher dimensional space-time. The Details of the higher-dimensional embedding of the theory will be discussed
in a separate paper \cite{large volume}.

\vspace{0.5cm}
\begin{center}{\bf ACKNOWLEDGMENTS}\end{center}
The current material was presented in the international Workshop on e-Science for Physics 2008 on September $8-9$, 2008, at Daejeon Convention Center, Daejeon, Korea. The content is based on the original research paper~\cite{Park and Yamaguchi} by the current author in collaboration with Satoshi Yamaguchi. The material has been presented at another place as well~\cite{Park:SUSY08}.



%
\newpage

\end{document}